\documentclass[aps,prb,superscriptaddress,twocolumn,amsmath,amssymb,  floatfix,  longbibliography]{revtex4-2} 
\usepackage{graphicx}
\usepackage{svg}
\usepackage{dcolumn}
\usepackage{bm}
\usepackage[mathlines]{lineno}

\usepackage{amsmath,amssymb}
\usepackage{graphicx}
\usepackage{booktabs}
\usepackage{float}
\usepackage{microtype}
\begin{document}

\title{Transport regimes in tunnel ferromagnetic Josephson junctions}
\author{R. Satariano} 
\affiliation{Department of Physics, University of Naples Federico II, 80125 Naples, Italy}
\author{A.A. Golubov} 
\affiliation{Moscow Institute of Physics and Technology, Dolgoprudny, Moscow Region 141701, Russia}
\affiliation{HSE University, 101000 Moscow, Russia}
\author{E. Raja} 
\affiliation{Department of Physics, University of Naples Federico II, 80125 Naples, Italy}
\author{R. Ferraiuolo} 
\altaffiliation[Present address: ] {QuantWare, Elektronicaweg 10, Delft, 2628 XG, Netherlands}
\affiliation{Department of Physics, University of Naples Federico II, 80125 Naples, Italy}
\author{F. Calloni} 
\affiliation{Department of Physics, University of Naples Federico II, 80125 Naples, Italy}
\author{H. G. Ahmad} 
\affiliation{Department of Physics, University of Naples Federico II, 80125 Naples, Italy}
\author{A. Vettoliere} 
\affiliation{Consiglio Nazionale delle Ricerche—ISASI, Via Campi Flegrei 34, I-80078 Pozzuoli, Italy}
\author{N. Poccia} 
\affiliation{Department of Physics, University of Naples Federico II, 80125 Naples, Italy}
\affiliation{Leibniz Institute for Solid State and Materials Research Dresden (IFW
 Dresden), 01069 Dresden, Germany }
\author{L. Parlato} 
\affiliation{Department of Physics, University of Naples Federico II, 80125 Naples, Italy}
\author{G. Ausanio} 
\affiliation{Department of Physics, University of Naples Federico II, 80125 Naples, Italy}
\author{C. Granata} 
\affiliation{Consiglio Nazionale delle Ricerche—ISASI, Via Campi Flegrei 34, I-80078 Pozzuoli, Italy}
\author{G. P. Pepe} 
\affiliation{Department of Physics, University of Naples Federico II, 80125 Naples, Italy}
\author{M. Yu. Kupriyanov} 
\affiliation{Skobeltsyn Institute of Nuclear Physics, Lomonosov Moscow State University, Moscow 119991, Russia}
\author{F. Tafuri} 
\affiliation{Department of Physics, University of Naples Federico II, 80125 Naples, Italy}
\author{D. Montemurro} 
\affiliation{Department of Physics, University of Naples Federico II, 80125 Naples, Italy}
\author{D. Massarotti} 
\affiliation{Department of Electrical Engineering and Information Technology, University of Naples Federico II, Naples I-80126, Italy}
\date{\today}

\begin{abstract}

We present a systematic study of the temperature-dependent transport properties of Nb- and Al-based superconductor--insulator--thin superconductor--ferromagnet--superconductor (SIsFS) Josephson junctions. By varying the superconducting material and the thickness of the intermediate superconducting layer, we identify three distinct transport regimes: a tunnel-dominated SIS-like regime, a temperature-driven crossover, and a complex weak-link regime. A unified Usadel-based description connects these regimes through the self-consistent pair potential of the intermediate s-layer.  While for Al-based devices an SIS-like transport regime is observed, a self-consistent calculation of the s/F/S proximity electrode, combined with perturbative tunnel coupling to the outer electrode, accounts for Nb devices with different thickness of the superconducting interlayer. The calculation places the 10-nm-interlayer sample in a proximity-supported, strongly suppressed \textit{s }state, close to the SInFS limit in which the interlayer is effectively driven normal (\textit{n}). A reduced nonlinear long-junction description captures the same limiting physics. The evolution of $I_\mathrm{c}R_\mathrm{N}(T)$ thereby provides a direct experimental signature of the three transport regimes. Beyond the fundamental interest, the results establish practical design guidelines: Nb junctions with sufficiently thick $s$ interlayers are well suited for high-speed cryogenic memory elements and tunable superconducting parametric amplifiers, while Al junctions are natural candidates for coherent superconducting quantum circuits.
\end{abstract}

\maketitle
\begin{figure*} [tb]    

\includegraphics[width = 17 cm]{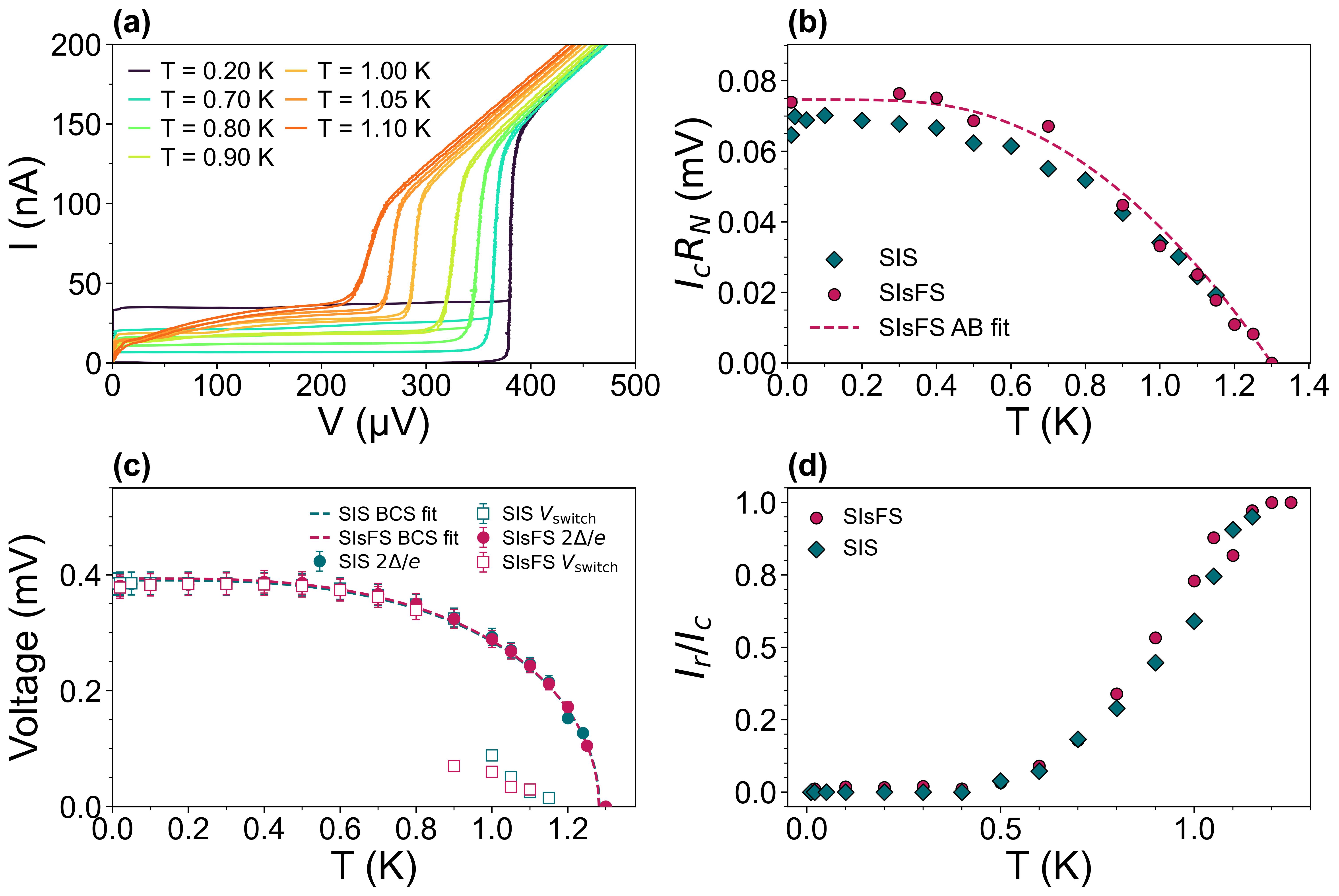}
\caption{Temperature evolution of the transport properties of circular Al-based superconductor--insulator--thin superconductor--ferromagnet--superconductor (SIsFS) Josephson junctions with a radius of 2~$\mu$m and a Py ferromagnetic layer [Al (200 nm)/AlO$_\mathrm{x}$ (2 nm)/Al (30 nm)/Py (3 nm)/Al (350 nm)], compared with reference SIsS junctions of the same geometry.
(a) Current--voltage characteristics of the Al-based SIsFS junction measured at
selected temperatures, as depicted in the legend.
(b) Critical current--normal state resistance product $I_\mathrm{c}R_\mathrm{N}$ as a function of
temperature for the SIsFS and SIsS junctions.
(c) Gap voltage $2\Delta/e$ (filled symbols) and switching voltage $V_{\mathrm{switch}}$ (open symbols) as a function of temperature for both devices.
(d) Retrapping-to-critical-current ratio $I_\mathrm{r}/I_\mathrm{c}$ as a function of temperature
for the SIsFS and SIsS junctions. In panels (b)--(d), the SIsFS and SIsS data are shown by magenta and teal dots, respectively.}
 \label{fig:al}
\end{figure*}
\section{INTRODUCTION} 
A ferromagnetic barrier in a Josephson junction (JJ) provides a versatile platform where superconductivity and ferromagnetism coexist, giving rise to a variety of emergent phenomena at superconductor/ferromagnet (S/F) interfaces \cite{GOLUBOV2004, BERGERET2005_REVIEW,Budzin_review_2005}, such as 0--$\pi$ phase transitions \cite{Ryazanov2001,kontos2002,Sellier2003,Frolov2004} and the generation of spin-triplet superconductivity \cite{Robinson_science_2010,Banerjee2014_bis,Banerjee2014,Singh2015,Martinez2016,Kapran2021}. This rich phenomenology  has established superconductor/ ferromagnet/ superconductor (SFS) junctions as a key building block for superconducting devices. In particular, the ability to engineer a $\pi$-shift has enabled the realization of passive phase shifters for superconducting digital and quantum circuits \cite{feofanov_pi_shifter,Birge2024,Kim2024}, whereas spin-triplet correlations provide the foundation for long-range spin transport and dissipationless spin manipulation in superconducting spintronics \cite{LINDER2015,ECHRING2015}. 

At the same time, the magnetic degree of freedom of ferromagnetic JJs has inspired the development of superconducting memory elements, in which distinct magnetic configurations correspond to different critical current values that encode logic states \cite{Soloviev2017,Birge2024,Mitrovic2025}. The memory properties of ferromagnetic JJs were first demonstrated in SFS junctions containing a single soft PdFe layer, where the remanent magnetization of the barrier alone sets the junction between two well-separated critical current levels \cite{BOLGINOV2012}.  Building on this result, several approaches have subsequently been proposed for implementing magnetic Josephson memories, including $\varphi$-junctions \cite{Goldobin_Phi_memory,Guarcello2020}, magnetic junctions based on pseudo-spin valves \cite{Banerjee2014,Baek2014_spin_valve, Qader2014_small_magnetic_fields_switching,Niedzielski2018} and multi-terminal Josephson devices \cite{Nevirkovets_multi_terminal}. However, the relatively low critical current--normal state resistance product $I_\mathrm{c}R_\mathrm{N}$ of conventional metallic SFS junctions, of the order of $\mu$V, limits their switching frequency and hinders compatibility with high-speed superconducting electronics \cite{Soloviev2017}.  In contrast, superconducting digital electronics, such as Single Flux Quantum (SFQ) circuits, predominantly rely on tunnel superconductor/ insulator/ superconductor (SIS) junctions with significantly larger $I_\mathrm{c}R_\mathrm{N}$ products of the order of mV, enabling switching frequencies in the hundreds-of-gigahertz range \cite{Soloviev2017, Mitrovic2025}. This limitation can be overcome by introducing  an insulating barrier together with an additional thin superconducting $s$ interlayer,  resulting in a hybrid SIsFS structure \cite{LARKIN2012,Vernik2013,CARUSO2018,karelina21_rectangular_SIsFS}. Such devices combine the high characteristic voltage of SIS tunnel junctions with the magnetic functionality of SFS weak-links, making them attractive candidates for superconducting memory technologies  \cite{Caruso2018_IEEE,JAP,Satariano_2024}.  High $I_cR_N$ is equally critical for twisted cuprate Josephson junctions
 proposed as key elements in new quantum hardwares \cite{Confalone_2025_ADT}, where membrane-based circuit technologies for 2D superconductors \cite{Saggau2023,confalone2026simplifiedsiliconnitridenanomembrane} preserve the Josephson coupling upon integration \cite{Confalone2025_SMALL}. Beyond cryogenic memories, the unique combination of magnetic control and low quasiparticle dissipation also makes SIsFS junctions attractive building blocks for coherent superconducting quantum circuits  \cite{ahmad2024,Satariano2025}. A prominent example is the ferro-transmon architecture, where the magnetic state of the junction provides a nonvolatile tuning knob for the Josephson energy and, consequently, for the qubit frequency control \cite{Halima_FERROTRASMONE, Massarotti2023Ferrotransmon}. More generally, the intrinsically digital switching capability of magnetic JJs opens the possibility of implementing qubit control using energy-efficient cryogenic digital electronics, offering a promising pathway toward scalable superconducting quantum processors \cite{McDermott_2018,Jordan26}. 

Apart from their technological relevance, SIsFS junctions represent a particularly rich physical system because of the coupling between the tunnel barrier and the magnetic weak-link mediated by a thin superconducting interlayer. The superconducting state of the intermediate $s$ layer is strongly affected by proximity-induced pair breaking and can evolve between qualitatively different regimes depending on its thickness, the properties of the adjacent ferromagnetic layer, and temperature. Theoretical studies have predicted a variety of nontrivial phenomena, including modifications of the current--phase relation and transitions between distinct transport regimes associated with the suppression or preservation of superconductivity in the $s$ layer \cite{Bakurskiy2013_APL,Bakurskiy2013_PRB,TAFURI2019}.  In particular, the superconducting state of the intermediate $s$ layer determines whether the junction behaves as a tunnel SIS device weakly perturbed by the ferromagnetic barrier or as a complex weak-link. More recently, related multilayer spin-valve structures have shown that the superconducting state of a thin $s$ layer can act as a trigger, enabling large magnetically controlled changes of the supercurrent \cite{Neilo2024_APL_trigger,Neilo2024_Nanomaterials_trigger}. As a consequence, SIsFS junctions provide a unique platform in which superconducting transport can be tuned through the interplay between superconducting and magnetic correlations, with temperature acting as an effective control parameter. 
In this work, we present a systematic experimental investigation of Nb- and Al-based SIsFS JJs by varying the thickness of the intermediate superconducting layer and following the evolution of their transport properties as a function of temperature down to 10 mK. The temperature dependence of the Josephson critical current emerges as the hallmark of different transport regimes, thus providing a means  to investigate the complex dynamics of SIsFS devices. Although $I_\mathrm{c}R_\mathrm{N}(T)$ does not determine every microscopic parameter uniquely, its shape is exceptionally sensitive to the superconducting state of the intermediate layer. In the present devices it distinguishes a robust condensate, a proximity-driven crossover, and a strongly suppressed-$s$ weak-link. The resulting picture obtained within a unified
Usadel-based description is corroborated by distinctive features of the temperature dependence of the current--voltage (I--V) characteristics.

 For Nb-based devices with sufficiently thick \textit{s}-layers, we identify a crossover from a tunnel-dominated regime, in which the junction behaves as a series combination of SIs and sFS elements, to a
proximity-driven weak-link regime as superconductivity in the intermediate
layer becomes progressively suppressed. Correspondingly, the temperature
dependence evolves from a tunnel-like behavior consistent with the
Ambegaokar-Baratoff relation \cite{AB_theory,BARONE1982} to an exponential decay of the
$I_\mathrm{c}R_\mathrm{N}$ product, characteristic of weak-link transport. By contrast,
junctions with thinner $s$ layers remain in the SIsFS weak-link regime across
the entire measured range down to 10 mK, while Al-based junctions remain
tunnel-like over the entire investigated temperature range. This comparison
demonstrates the decisive role of the superconducting materials intermediate-layer thickness and of the transparency of the S/F interface in determining the operating regime of hybrid SIsFS junctions.  
It points to Nb-based devices as natural candidates for
high-speed cryogenic memory elements \cite{Soloviev2017} and  tunable superconducting parametric amplifiers \cite{JPA_patent}, and
Al-based devices for coherent superconducting quantum circuits \cite{Halima_FERROTRASMONE,Satariano2025, FerromagneticPatent}.

\section{Fabrication and experimental methods}
Three sets of SIsFS Josephson junctions were fabricated and investigated in this work, differing in the superconducting electrode material and the thickness of the intermediate superconducting layer. All devices were fabricated starting from either a Nb/AlO$_\mathrm{x}$/Nb or Al/AlO$_\mathrm{x}$/Al (SIS) trilayer deposited by dc magnetron sputtering in an ultrahigh-vacuum system. The base and intermediate superconducting electrodes were deposited at a rate of 1.2~nm/s, while the Al layer used to form the tunnel barrier was deposited at a reduced rate of 0.7~nm/s and subsequently exposed to dry oxygen for 1~h to form the AlO$_\mathrm{x}$ barrier \cite{JAP,vettoliere_nanomaterials}. The trilayer was patterned by optical lithography and lift-off, while the junction area was defined by selective anodization and electrically insulated by SiO$_2$ deposition \cite{JAP,vettoliere_nanomaterials}. Prior to ferromagnetic-layer deposition, the surface of the thin superconducting interlayer was cleaned by soft Ar ion milling to prepare the S/F interface. The ferromagnetic layer was subsequently deposited by magnetron sputtering at a rate of 0.7~nm/s using a lift-off process, followed by deposition of the top superconducting electrode by dc magnetron sputtering and lift-off, completing the SIsFS stack \cite{JAP,vettoliere_nanomaterials}.
Sample A consists of Al (200 nm)/AlO$_\mathrm{x}$ (2 nm)/Al (30 nm)/Py (3 nm)/Al (350 nm). Sample B consists of Nb (200 nm)/AlO$_\mathrm{x}$ (2 nm)/Nb (30 nm)/Py (3 nm)/Nb (350 nm). 
Sample C consists of Nb (200 nm)/AlO$_\mathrm{x}$ (2 nm)/Nb (10 nm)/Py (3 nm)/Nb (350 nm). Reference SIsS junctions were fabricated on the same wafer by excluding the ferromagnetic-layer deposition step.

Transport measurements were performed in an Oxford Instruments Triton dry dilution refrigerator. The samples were thermally anchored to the mixing chamber and connected through custom low-noise filtered wiring thermalized at various temperature stages of the cryostat \cite{Ahmad_22}. The junctions were current biased using a low-frequency ($\sim$1.1~Hz) current ramp generated by a waveform generator connected in series with a shunt resistor. The voltage across the junction was measured using a battery-powered differential amplifier.

\section{EXPERIMENTAL RESULTS} 
The transport properties of SIsFS Josephson junctions strongly depend on the thickness $d_\mathrm{s}$ of the intermediate superconducting layer relative to the critical thickness $d_\mathrm{sc}$, defined as the minimum thickness for which superconductivity survives in an s/F bilayer at a given temperature \cite{Bakurskiy2013_APL,Bakurskiy2013_PRB}. If $d_\mathrm{s}$ is sufficiently larger than $d_\mathrm{sc}$, the pair potential in the s layer $\Delta_\mathrm{s}$ is essentially equal to its bulk value and the SIsFS JJ can be regarded as a series connection of a tunnel SIs junction and a metallic sFS junction. For small $d_\mathrm{F}$ and sufficiently transparent $s/F$ and $F/S$ interfaces, the critical current of the sFS side is expected to be much larger than that of the tunnel SIs junction, i.e., $I_{1c} \ll I_{\mathrm{2c}}$, where $I_{\mathrm{1c}}$ and $I_{\mathrm{\mathrm{2c}}}$ are the critical currents of the SIs and sFS sides, respectively. As a result, the transport properties of the complete SIsFS junction are governed by the SIs part, and the critical current--normal resistance product $I_\mathrm{c}R_\mathrm{N}$ can reach its maximum value corresponding to a standard SIS JJ \cite{Bakurskiy2013_APL, Bakurskiy2013_PRB}. The junction parameters of all samples, measured at $T = 10$ mK, are summarized in Table \ref{Tabella1}. 

Sample A clearly falls into this regime. Figure~\ref{fig:al}(a) shows the current--voltage (I--V) characteristics measured at different temperatures. The electrodynamic properties of the SIsFS junction are indistinguishable from those of the corresponding reference SIsS junction fabricated on the same wafer \cite{Vettoliere_APL,ahmad2024}. In particular, the temperature dependences of the characteristic voltage $I_\mathrm{c}R_\mathrm{N}$ [Fig.~\ref{fig:al}(b)], the superconducting gap voltage $2\Delta/e$ [Fig.~\ref{fig:al}(c)], and the retrapping-to-critical-current ratio $I_\mathrm{r}/I_\mathrm{c}$ [Fig.~\ref{fig:al}(d)] coincide with those measured in the reference SIsS device. These results demonstrate that the insertion of the ferromagnetic layer with this choice of materials and thickness does not measurably modify the transport properties. Two mechanisms may contribute. First, the superconducting energy gap of thin Al films is known to increase as the film thickness decreases \cite{universal_scaling_critical_temperature,Critical_temperature_Al_thin_film,marchigiani_Al}. For film thicknesses larger than approximately 3~nm, the gap follows the phenomenological relation $\Delta(d)=\Delta_{\mathrm{bulk}}+ad^{-1}$, where $\Delta_{\mathrm{bulk}}\simeq180\ \mu\mathrm{eV}$ and $a$ depends on the deposition conditions \cite{marchigiani_Al}. Since the Al interlayer in Sample A is thinner than the electrodes, its superconducting gap is expected to be slightly larger than that of the electrodes. The intermediate layer can therefore remain fully superconducting throughout the investigated range and satisfy $I_{\mathrm{1c}}\ll I_{\mathrm{2c}}$, leaving the tunnel SIs element to control the overall transport \cite{Satariano2025}. Second, a native or processing-induced AlO$_{\mathrm{x}}$ layer at the Al/Py interface \cite{EuS_PRB,EuS_nature_communication} may transmit exchange correlations only weakly.  Within the proximity framework, a reduced effective interface transparency has to be taken into account to suppress the influence of Py on the Al interlayer.

\begin{figure}[h]
\includegraphics[width = 6.5 cm]
{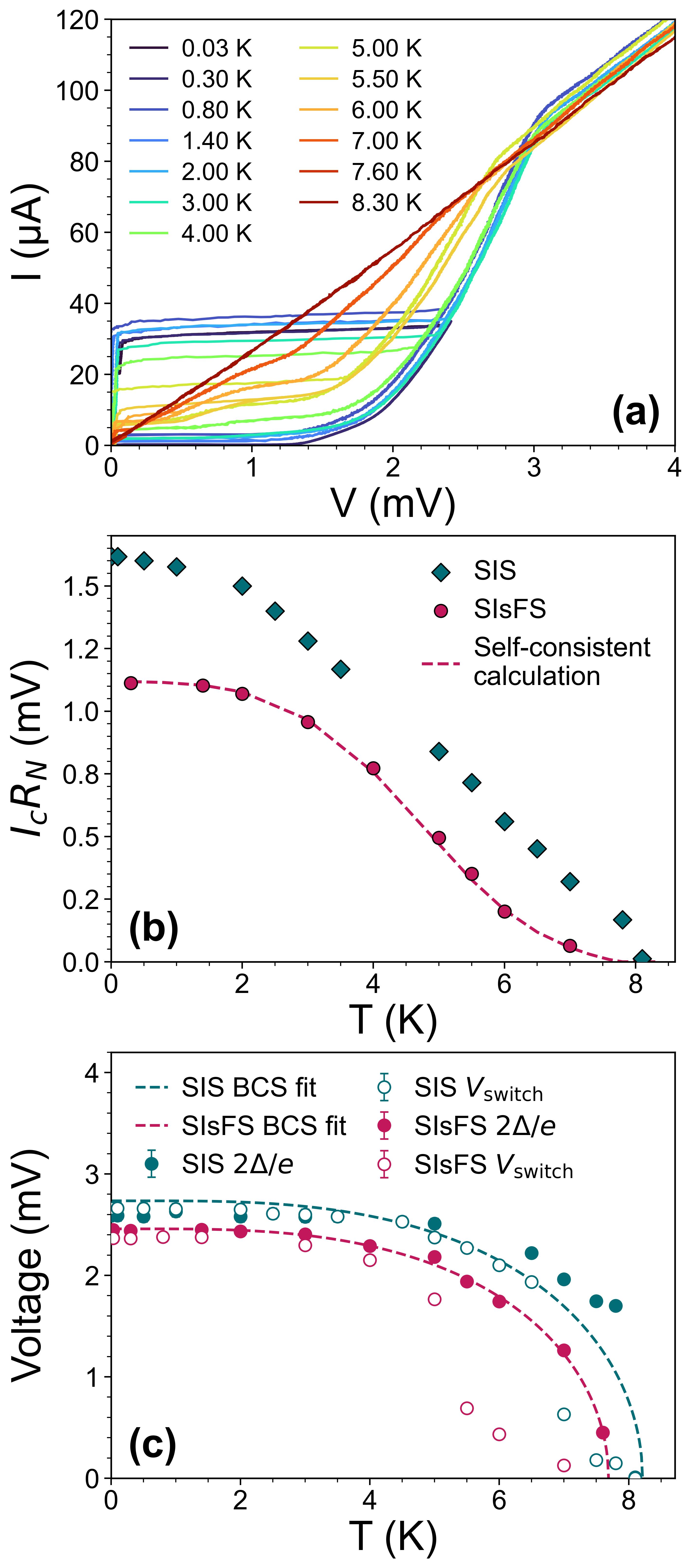}
\caption{Temperature evolution of the transport properties of a square Nb-based SIsFS Josephson junction Nb (200 nm)/AlO$_\mathrm{x}$ (2 nm)/Nb (30 nm)/Py (3 nm)/Nb (350 nm) with a lateral size of 10~$\mu$m, compared with reference SIsS junctions. (a) Current--voltage characteristics of the SIsFS junction measured at selected
temperatures, as depicted in the panel. 
(b) Characteristic voltage $I_\mathrm{c}R_\mathrm{N}$ as a function of temperature for SIsFS and SIsS junctions. The magenta line is the self-consistent result for
Sample B; model parameters and spatial profiles are given in
the Supplemental Material. (c) Gap voltage $2\Delta/e$ (filled symbols) and switching voltage
$V_{\mathrm{switch}}$ (open symbols) as a function of temperature for both devices; the device identity is indicated in the legend.
In panels (b)--(c), the SIsFS and SIsS data are shown by magenta and teal dots, respectively.
} \label{fig:IVT-Nb-30-nm}
\end{figure}

Unlike Sample A, the transport properties of Sample B exhibit a clear temperature dependence, reflecting the progressive modification of the $s$ interlayer by proximity effects. The corresponding I--V characteristics are reported in Fig.~\ref{fig:IVT-Nb-30-nm}(a). An isolated 30-nm Nb film has a critical-temperature scale close to 6~K~\cite{cirillo_Tc}; we therefore denote $T^*\simeq6$~K as the intrinsic pairing scale of the intermediate Nb layer.   
At low temperature, the interlayer remains sufficiently robust that the phase drop is localized mainly across the insulating barrier and the device retains predominantly tunnel-like transport. Nevertheless, the ferromagnetic layer already reduces $I_\mathrm{c}R_\mathrm{N}$: at $T=10$~mK, $I_\mathrm{c}R_\mathrm{N}=1.1$~mV for the 10-$\mu$m square junction, approximately 30\% below the reference SIsS value of 1.6~mV. With increasing temperature, proximity-induced pair breaking from Py
progressively suppresses the pair potential near the $s/F$ interface. In the complete SIsFS structure, however,  $T^*$ sets a crossover scale rather than a transition of the complete structure: above it the \textit{s} interlayer no longer sustains an autonomous condensate, but the right Nb electrode remains superconducting up to approximately 8~K and transmits anomalous correlations through Py into the intermediate layer.  These correlations sustain the high-temperature current tail above the $T^*$ scale, while Py simultaneously produces pair breaking. The measured tail in Fig.~\ref{fig:IVT-Nb-30-nm}(b) therefore reflects the competition between reverse-proximity suppression and proximity-induced pairing from the right superconducting electrode $S_\mathrm{R}$. The progressive deviation of the temperature dependence of the switching voltage $V_{\mathrm{switch}}$ from the SIsS value as $T$ approaches $T^*$ [Fig.~\ref{fig:IVT-Nb-30-nm}(c)] provides an independent signature of the same crossover.

\begin{figure*}[tb]
\includegraphics[width = 18cm]{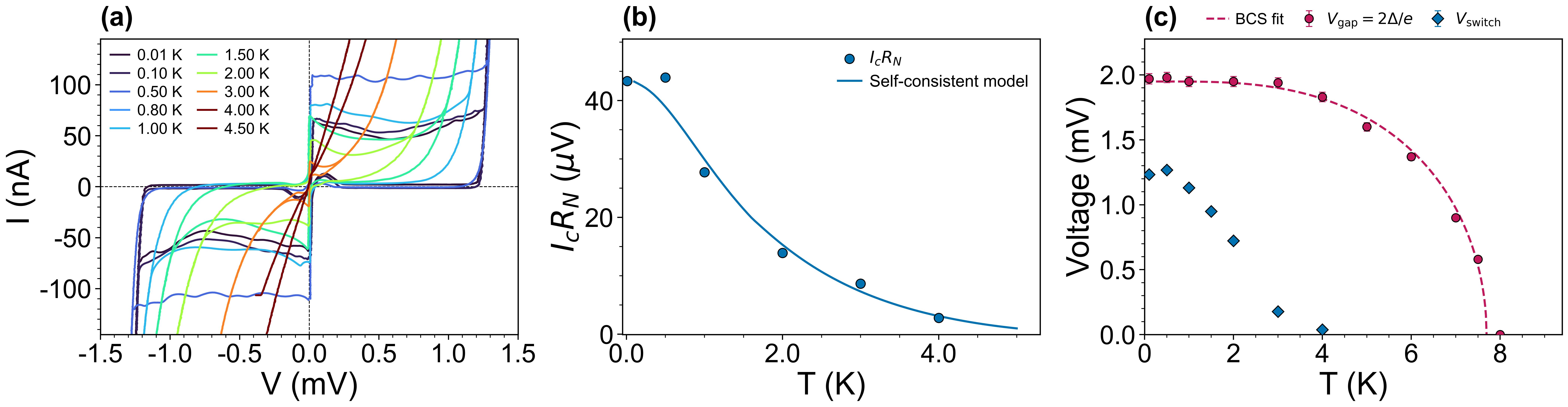}
\caption{Temperature evolution of the transport properties of a circular
Nb-based SIsFS Josephson junction with $d_\mathrm{s} = 10$~nm, a Py ferromagnetic
barrier, and a diameter of 2~$\mu$m.
(a) Current--voltage characteristics measured at selected temperatures, as
depicted in the legend.
(b) Characteristic voltage $I_\mathrm{c}R_\mathrm{N}$ as a function of temperature. Blue circles are experimental data and the solid blue curve is the self-consistent result for the nominal 10-nm interlayer, evaluated from $0.1$ to $5$ K. The calculation captures the strongly suppressed-$s$ proximity regime; model parameters and spatial profiles are given in the Supplemental Material.
(c) Gap voltage $V_{\mathrm{gap}}=2\Delta_0/e$ (magenta circles) and switching voltage
$V_{\mathrm{switch}}$ (green diamonds) as a function of temperature. The dashed magenta line is a normalized BCS fit with $V_{\mathrm{gap}}(0)=1.94$~mV and $T_c^{\mathrm{gap}}=7.70$~K.}
 \label{fig:IVT-Nb-10-nm}
\end{figure*}

For Sample C, the most relevant feature of the data reported in Fig.~\ref{fig:IVT-Nb-10-nm} is the strong suppression of both the critical current density and the characteristic voltage $I_\mathrm{c}R_\mathrm{N}$, by factors of roughly 20--50 relative to the reference SIsS junction and Sample B [compare Figs.~\ref{fig:IVT-Nb-10-nm}(b) and~\ref{fig:IVT-Nb-30-nm}(b)]. This behavior indicates a markedly different regime from Samples A and B. Here, the intermediate-layer thickness is comparable to the superconducting coherence length ($d_\mathrm{s}/\xi_\mathrm{s} \sim 1$), so proximity- and exchange-induced pair breaking act across essentially the entire interlayer. As shown by the self-consistent calculation below, the pair potential in $s$ is strongly reduced and transport is governed by the composite $s/F/S$ proximity electrode. No critical current is observed above approximately 4.5~K, while the measured gap voltage follows the normalized BCS temperature dependence with $V_{\mathrm{gap}}(0)=1.94$~mV and $T_c^{\mathrm{gap}}=7.70$~K [Fig.~\ref{fig:IVT-Nb-10-nm}(c)]. This scale is close to the 8-K transition of the Nb electrodes, showing that their local pairing amplitude survives even though the supercurrent is strongly suppressed. The switching voltage $V_{\mathrm{switch}}$ remains below $2\Delta/e$ over the entire investigated temperature [Fig.~\ref{fig:IVT-Nb-10-nm}(c)]. Together with the early disappearance of hysteresis in the I--V characteristics, as shown by the temperature dependence of the retrapping-to-critical-current ratio  $I_\mathrm{r}/I_\mathrm{c}$ in Fig. \ref{fig:comparison-nb}, this separation provides independent qualitative evidence that Sample C has departed from conventional tunnel-junction dynamics.
\begin{table}[h]
\centering
\small
\setlength{\tabcolsep}{3.5pt}
\caption{Parameters of the samples measured at $T=10$~mK. Here $A$ is the junction area, $j_\mathrm{c}$ is the critical current density, $I_\mathrm{c}R_\mathrm{N}$ is the characteristic voltage, $R_\mathrm{N}A$ is the resistance--area product, and $2\Delta/e$ is the gap voltage.}
\label{Tabella1}
\begin{tabular}{c c c c c c}
\hline
 & $A$ & $j_\mathrm{c}$ & $I_\mathrm{c}R_\mathrm{N}$& $R_\mathrm{N}A$&  2$\Delta/e$ \\
Sample & ($\mu$m$^2$) & (A/cm$^2$) & (mV) & (k$\Omega\,\mu$m$^2$) & (mV) \\
\hline
SIsS Al& 13  & 0.3& 0.07& 23& 0.4 \\
SIsFS Al& 13  & 0.3& 0.07& 23& 0.4 \\
SIsS Nb& 13  & 160 & 1.6  & 1.0   & 2.6 \\
SIsFS B  & 100 & 30  & 1.1  & 3.6& 2.5 \\
SIsFS C  & 3   & 3.5& 0.05& 1.3& 1.9  \\
\hline
\end{tabular}
\end{table}

\section{THEORETICAL MODEL AND DISCUSSION}
The three regimes can be organized within one dirty-limit Usadel framework by following how the pair potential in the intermediate layer evolves from a robust superconducting value to a strongly suppressed one. The tunnel barrier has the largest resistance and is treated perturbatively. The right proximity electrode $s/F/S_\mathrm{R}$ is solved self-consistently, while the left superconducting electrode enters through its bulk spectral angle. We use the $\theta$ parametrization
\begin{equation}
F=\sin\theta,
\qquad
G=\cos\theta,
\end{equation}
which automatically satisfies $G^2+F^2=1$. For material $j$, let $y_j=x/\xi_{D,j}$, where $\xi_{D,j}=\sqrt{\hbar D_j/(2\pi k_BT_c)}$, and define $\Omega_n=\omega_n/(\pi k_BT_c)=(2n+1)T/T_c$ and $\delta_j=\Delta_j/(\pi k_BT_c)$. Primes below denote differentiation with respect to the local coordinate $y_j$. In the $s$ and $F$ layers,
\begin{align}
\theta_s''-\Omega_n\sin\theta_s
+\delta_s(y_s)\cos\theta_s &=0,
\label{eq:usadel_s}\\
\theta_F''-\left(\Omega_n+ih\right)\sin\theta_F &=0,
\label{eq:usadel_f}
\end{align}
where $h=H/(\pi k_BT_c)$ is the normalized exchange energy. The imaginary shift in the ferromagnet produces the oscillatory anomalous amplitude responsible for $0$--$\pi$ behavior. In the fits below we denote this parameter by $h^{\rm eff}$ to emphasize that, within a one-dimensional diffusive description of a processed ultrathin Py region, it is an effective exchange and pair-breaking scale rather than a direct determination of the intrinsic bulk-Py exchange field. The right electrode $S_\mathrm{R}$ obeys the superconducting form of Eq.~\eqref{eq:usadel_s}, with $\delta_\mathrm{s}$ replaced by $\delta_\mathrm{R}$. In both superconducting regions the pair potential is determined from the local self-consistency equation; for the intermediate layer,
\begin{equation}
\delta_s(y_s)\ln\frac{T}{T_{c,s}}
+2\frac{T}{T_c}\sum_{n\geq0}
\left[
\frac{\delta_s(y_s)}{\Omega_n}
-\operatorname{Re}\sin\theta_s(\Omega_n,y_s)
\right]=0,
\label{eq:selfcons}
\end{equation}
with the corresponding equation in $S_\mathrm{R}$ obtained by setting the intrinsic transition temperature equal to the electrode value $T_c$. Here $T_{c,s}$ is the intrinsic pairing scale entering the self-consistency equation, not the transition temperature of the proximity-coupled layer. Equation~\eqref{eq:selfcons} allows the same formulation to describe a nearly bulk-like $s$ layer (Sample A and Sample B at low temperature) and a strongly suppressed one (Sample C).

At the $s/F$ interface we impose the Kupriyanov--Lukichev boundary conditions~\cite{Kurpianov1988_SN}
\begin{align}
\partial_{y_s}\theta_s
&=\gamma_{sF}\partial_{y_F}\theta_F,
\\
\gamma_{B,sF}\partial_{y_F}\theta_F
&=\sin(\theta_F-\theta_s),
\label{eq:KL}
\end{align}
and at $F/S_\mathrm{R}$ the corresponding conditions are
\begin{align*}
\partial_{y_R}\theta_R
&=\gamma_{FS}\partial_{y_F}\theta_F,\\
\gamma_{B,FS}\partial_{y_F}\theta_F
&=\sin(\theta_R-\theta_F).
\end{align*}
Here $\gamma_{sF}=\rho_s\xi_{D,s}/(\rho_F\xi_{D,F})$ and $\gamma_{FS}=\rho_R\xi_{D,R}/(\rho_F\xi_{D,F})$, while $\gamma_{B,sF}=R_{B,sF}/(\rho_F\xi_{D,F})$ and $\gamma_{B,FS}=R_{B,FS}/(\rho_F\xi_{D,F})$; $R_{B,sF}$ and $R_{B,FS}$ are the corresponding interface resistances per unit area. At the high-resistance $I/s$ boundary the zeroth-order proximity problem is reflecting, $\partial_{y_s}\theta_s=0$. The right electrode is taken sufficiently thick for the order parameter to recover its bulk value. At its free outer surface the absence of spectral and condensate flux requires
\begin{equation*}
\left.\partial_x\theta_R\right|_{x_{\rm out}}=0,
\qquad
\left.\partial_x\Delta_R\right|_{x_{\rm out}}=0.
\end{equation*}
The common thermal diffusion-length convention is
\begin{equation}
\xi_{D,j}=\sqrt{\frac{\hbar D_j}{2\pi k_BT_c}}.
\end{equation}

Because $R_\mathrm{N}$ is dominated by the AlO$_x$ barrier (Table~\ref{Tabella1}), the signed first Josephson harmonic is evaluated at the tunnel side, $x=0$, as
\begin{equation}
I_1R_\mathrm{N}\simeq \zeta_I\frac{2\pi k_BT}{e}
\operatorname{Re}\sum_{n\geq0}
\sin\theta_L(\Omega_n)\sin\theta_s(\Omega_n,0),
\label{eq:tunnel_current}
\end{equation}
where $\theta_L=\arctan[\Delta_L(T)/\omega_n]$ is the bulk spectral angle of the left electrode and $\zeta_I$ is a tunnel-side spectral factor ($\zeta_I=1$ in the ideal limit). The measured critical current is $I_\mathrm{c}=|I_1|$. Interface effects enter Eq.~\eqref{eq:tunnel_current} through the spectral amplitudes, not through an additional series resistance.

For Sample A, the thickness-enhanced gap of the thin Al interlayer and weak effective exchange coupling across the processed Al/Py interface place the device in the limit $\Delta_s\simeq\Delta_{\rm Al}$. The condition $I_{\mathrm{1c}}\ll I_{\mathrm{2c}}$ is then satisfied, and Eq.~\eqref{eq:tunnel_current} reduces to the Ambegaokar--Baratoff dependence observed experimentally, with no measurable signature of the ferromagnetic layer [Fig.~\ref{fig:al}(b)].

To compare the two Nb devices on a common footing, we use the same electrode and interlayer pairing scales, coherence-length normalization, and metallic-interface mismatch. Within this parameter set, the self-consistent calculation reproduces the measured  $I_\mathrm{c}R_\mathrm{N} (T)$ of both devices, which differ by more than an order of magnitude and qualitatively in shape.  The common value $\xi_\mathrm{s}=10.7$~nm converts the nominal 30- and 10-nm interlayers to $d_\mathrm{s}/\xi_\mathrm{s}=2.80$ and $0.935$, respectively. A small sample-dependent variation of the overall tunnel normalization is allowed; the remaining parameters are given in the Supplemental Material.

For Sample B, the self-consistent solution reproduces both the low-temperature scale and the measured $5.5$--$8$ K tail. Above the bare interlayer scale $T^*\simeq6$~K, $\Delta_s$ is sustained by correlations transmitted from $S_\mathrm{R}$ through Py. The ferromagnetic layer therefore plays a dual role: the reverse-proximity suppresses the adjacent condensate, while transmission from $S_\mathrm{R}$ supports the high-temperature tail. The spatial profiles and tunnel-side amplitudes are shown in Figs.~S1--S3 of the Supplemental Material.

For Sample C, the nominal 10-nm thickness gives $d_\mathrm{s}/\xi_\mathrm{s}=0.935$. The self-consistent solution yields an interlayer pair potential far below the outer-electrode scale while retaining a finite anomalous amplitude at the tunnel boundary (see Figs.~S4--S6 of the Supplemental Material). Sample C is therefore in a proximity-supported, strongly suppressed-$s$ regime close to the SInFS limit: the tunnel current couples $S_L$ to the composite $s/F/S_\mathrm{R}$ proximity electrode, rather than to a robust independent $s$ electrode. The Sample-C solution lies on the $\pi$ branch throughout the measured range. Since a conventional current--voltage characteristic measures $I_\mathrm{c}=|I_1|$, this sign is a model prediction; its verification requires a phase-sensitive interferometric measurement.

The two Nb samples are described using common superconducting parameters, whereas their distinct proximity regimes require sample-dependent effective Py propagation and metallic-interface parameters. In Sample C, reduced interface transparency weakens both exchange-induced pair breaking of the thin $s$ layer and transmission of correlations from $S_\mathrm{R}$, so the calculated response reflects a balance between these effects. Because the effective parameters are correlated, they should not be interpreted as unique microscopic determinations.

\begin{figure}[h]
\includegraphics[width=8 cm]{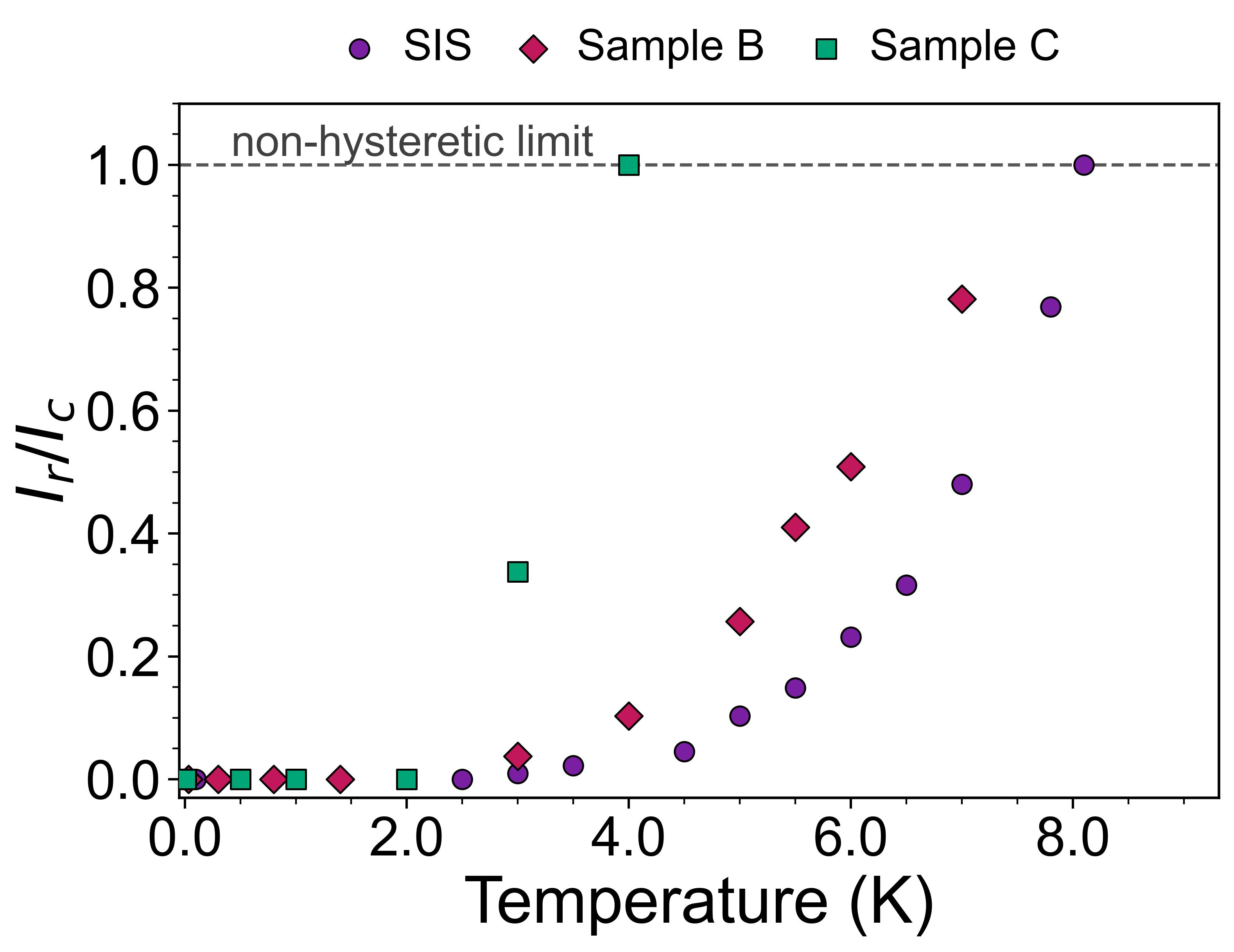}
\caption{Temperature dependence of the retrapping-to-critical-current ratio $I_\mathrm{r}/I_\mathrm{c}$ for the Nb-based SIsS reference and the SIsFS junctions with a 30-nm Nb interlayer (Sample B) and a 10-nm Nb interlayer (Sample C). Symbols are experimental data; thin connecting lines are guides to the eye. The dashed line at $I_\mathrm{r}/I_\mathrm{c}=1$ marks the non-hysteretic limit. The much earlier rise for Sample C shows that the 10-nm device loses resolvable hysteresis well below the corresponding temperatures for Sample B and the SIsS reference.}
\label{fig:comparison-nb}
\end{figure}

For a complementary analytical comparison with the suppressed-$s$ interpretation, we also consider the reduced nonlinear long-junction limit. When $\Delta_\mathrm{s}\simeq0$, the normal-like segment obeys
\begin{equation}
\theta''-\kappa_n^2\sin\theta=0, \qquad \kappa_n=\sqrt{\frac{2\omega_n}{\hbar D}},
\label{eq:nonlinear_usadel}
\end{equation}
and the long-segment solution can be written as the superposition of the source tail and its reflecting image,
\begin{equation}
\begin{aligned}
\tan\frac{\theta(x,\omega_n)}{4}
&\simeq q_n\left[e^{-\kappa_n(L-x)}+e^{-\kappa_n(L+x)}\right],\\
q_n&=\tan\frac{\theta_{b,n}}{4}.
\end{aligned}
\label{eq:tail_solution}
\end{equation}
Here $x=0$ is the reflecting tunnel boundary and $x=L$ is the source side. Equation~\eqref{eq:tail_solution} satisfies $\theta'(0)=0$ asymptotically and gives
\begin{equation}
\tan\frac{\theta(0,\omega_n)}{4}\simeq u_n,
\qquad u_n=2q_ne^{-\Lambda_n},
\label{eq:un_def}
\end{equation}
with $\Lambda_n = \kappa_N L$.
The exact trigonometric relation is
\begin{equation}
\sin\theta(0,\omega_n) = \frac{4u_n(1-u_n^2)}{(1+u_n^2)^2}.
\label{eq:sintheta0}
\end{equation}
Equations~\eqref{eq:tunnel_current} and~\eqref{eq:sintheta0} give the nonlinear long-junction tunnel expression
\begin{equation}
I_\mathrm{c}R_\mathrm{N} = \zeta_I \frac{2\pi k_BT}{e}\left|\operatorname{Re}\sum_{n=0}^{\infty}\sin\theta_L(\omega_n)\frac{4u_n(1-u_n^2)}{(1+u_n^2)^2}\right|.
\label{eq:IcRN_nonlinear_general}
\end{equation}
For the reduced Sample-C comparison we use
\begin{equation}
q_n=\eta_q\tan\frac{\theta_R(\omega_n)}{4},
\qquad
\Lambda_n=\sqrt{\frac{2\omega_n}{E_{\mathrm{Th}}^{\mathrm{eff}}}}.
\label{eq:effective_tail}
\end{equation}
Here $\theta_R(\omega_n)=\arctan[\Delta_R^{\rm bulk}(T)/\omega_n]$ is the bulk spectral angle of the right source electrode, and the reduction factor $\eta_q$ represents the non-rigid $F/S_\mathrm{R}$ source amplitude. The fitted $E_{\mathrm{Th}}^{\mathrm{eff}}$ is a composite spectral scale that contains propagation through the suppressed Nb layer, damping and phase rotation in Py, and finite-interface suppression. Related boundary-induced renormalization of the effective Thouless scale has been analysed explicitly for SN--N--NS bridges \cite{Bakurskiy2024_Thouless}. In the strong-ferromagnet limit the Py factor is approximately
\begin{equation*}
\exp\!\left[-(1+i)\frac{d_F}{\xi_{D,F}}\sqrt{\frac{h}{2}}\right],
\end{equation*}
and is nearly temperature independent; the dominant thermal dependence is therefore controlled by the suppressed 10-nm Nb interlayer.  The reduced model reproduces the Sample-C temperature dependence over $0.5$--$4$ K with $E_{\mathrm{Th}}^{\mathrm{eff}}\simeq70~\mu$eV.  This scale characterizes propagation through the composite proximity region rather than an ideal SNS link. The reduced model provides a complementary description of the strongly suppressed-$s$ limit and is not used to determine the self-consistent parameters; fit parameters and the comparison including the 10-mK point are given in the Supplemental Material.
For an ideal long diffusive SNS junction, the induced minigap is set by the Thouless scale \cite{Dubos2001}. In the present S--I--proximized structure, however, $E_{\mathrm{Th}}^{\mathrm{eff}}$ is a composite fit parameter and $V_{\mathrm{switch}}$ is a dynamical quantity affected by the tunnel barrier, damping, and the full spectral structure. We therefore do not identify $eV_{\mathrm{switch}}$ directly with an induced minigap. The reduced switching voltage and the thermal evolution of the retrapping-to-critical-current ratio $I_\mathrm{r}/I_\mathrm{c}$ nevertheless provide independent qualitative support for departure from the tunnel-dominated regime (Fig.~\ref{fig:comparison-nb}). At base temperature, Samples B and C have comparably small ratios. With increasing temperature, $I_\mathrm{r}/I_\mathrm{c}$ rises much earlier in the 10-nm device, reaching the non-hysteretic limit near 4~K, whereas the 30-nm device and the SIsS reference remain hysteretic to higher temperatures.

Beyond establishing the microscopic mechanisms governing transport in these hybrid junctions, these results translate into concrete design guidelines for SIsFS-based magnetic switching elements. In such applications, the interlayer thickness $d_s$ is typically chosen below the London penetration depth $\lambda_L$, so as to preserve the memory properties of standard SFS JJs \cite{LARKIN2012,karelina21_rectangular_SIsFS}. Our results show that this constraint is handled very
differently by the two superconducting materials studied here. For
Al-based junctions, $d_\mathrm{s}$ can be reduced toward this limit essentially
without penalty: the thickness-enhanced gap and weak effective coupling across the processed Al/Py interface keep the junction locked in the tunnel-dominated regime (Sample A). For Nb-based junctions, by contrast, a lower bound on $d_\mathrm{s}$ must be set more carefully: as $d_\mathrm{s}$ approaches the coherence length $\xi_\mathrm{s}$ (Sample C), proximity-induced pair breaking from the adjacent ferromagnet progressively suppresses the interlayer order parameter, driving the junction into a complex weak-link regime and degrading both the characteristic voltage $I_\mathrm{c}R_\mathrm{N}$ and, correspondingly, the switching speed.
The pair-breaking strength depends on the exchange field of the  ferromagnet and on the transparency of the $s$/F  interface: the stronger the ferromagnet, the thicker the $s$ interlayer required to keep the device on the tunnel-dominated side of the crossover. This trade-off becomes
particularly relevant for submicron devices aimed at high-density cryogenic memories, which require ferromagnetic materials with large exchange fields
such as Py in order to resolve two distinct critical-current states in the magnetic field pattern \cite{JAP,Satariano2025,FerromagneticPatent}. The crossover identified for Sample B ($d_\mathrm{s} \sim 3\,\xi_\mathrm{s}$) thus marks a
practical design boundary: below it, Nb-based SIsFS memory elements risk losing the high-$I_\mathrm{c}R_\mathrm{N}$ and high-speed advantage that motivates the SIsFS
architecture in the first place.  

\section{CONCLUSIONS}
We have presented a systematic experimental investigation of Nb- and Al-based SIsFS Josephson junctions, following their transport properties down to 10~mK while varying the material and thickness of the intermediate superconducting layer. Within one set of devices we identify three regimes previously predicted theoretically or accessed separately: a tunnel-dominated SIS-like regime (Sample A), a temperature-driven crossover (Sample B), and a complex weak-link regime (Sample C).
The Al-based device (Sample A) shows no measurable trace of the ferromagnetic barrier at any temperature, consistent with the combined effect of the thickness-enhanced superconducting gap of the thin Al interlayer and weak effective coupling across the processed Al/Py interface. The Nb-based device with a nominal 30-nm interlayer (Sample B) instead shows a $30\%$ reduction of the characteristic voltage at low temperature and a proximity-supported tail above the bare interlayer scale $T^*\simeq6$~K. The self-consistent calculation supports this interpretation: although Py suppresses the adjacent interlayer through reverse-proximity effect, it also transmits superconducting correlations from $S_\mathrm{R}$, so that a finite anomalous amplitude reaches the tunnel barrier up to the transition of the outer Nb electrodes. The Nb-based device with a 10-nm interlayer (Sample C) is in the weak-link regime across the measured range, with a characteristic voltage reduced by more than an order of magnitude and no measurable critical current above 4.5~K. Within the same intrinsic Nb scale and closely matched tunnel normalizations, the self-consistent $s/F/S_\mathrm{R}$ calculation combined with perturbative tunnel coupling describes both Nb devices and connects the crossover to the reduction of $d_s/\xi_s$ from $2.80$ to $0.935$. The reduced nonlinear long-junction limit provides a complementary description of Sample C in the suppressed-$s$ limit. 
The central result is the evolution from a relatively robust interlayer in Sample B to a proximity-supported, strongly suppressed-$s$ state in Sample C.
These results provide practical guidance for engineering SIsFS devices: the large characteristic voltage and
temperature stability of Nb-based junctions with sufficiently thick
interlayers make them well suited for high-speed cryogenic memory elements \cite{Soloviev2017} and parametric amplifiers \cite{JPA_patent}, whereas the robustness of Al-based junctions against the ferromagnetic barrier makes them natural candidates for coherent superconducting quantum circuits such as the ferro-transmon \cite{Halima_FERROTRASMONE, Satariano2025}.
\section*{Acknowledgements}
This work has been supported by the Pathfinder EIC 2023 project "FERROMON-Ferrotransmons and Ferrogatemons for Scalable Superconducting Quantum Computers" , Project SUPREME - SUPERCONDUCTING EUROPEAN QUANTUM PILOT LINE (HORIZON-JU-CHIPS-2025-QAC1-1-SGA) e and Project Q-SUD, CUP B69H26001120007, within the framework of the Italian PN RIC 2021-2027 Programme. R.S. acknowledges the project “MoQua–
Novel Magnetic Orders for superconducting Quantum Architectures”–Ricerca di Ateneo Linea C. A.G. and M.Yu.K. acknowledge support by the Russian Science Foundation project No 23-72-30004. N.P. acknowledges the partial  funding by the European Union (ERC-CoG, 3DCuT, 101124606), the Deutsche 
 Forschungsgemeinschaft (DFG, German Research Foundation): DFG 512734967,  DFG 452128813, DFG 539383397, DFG 572638824. 
\section*{Conflict of Interest} The authors have no conflicts to disclose. 
\section*{Data Availability Statement} The data that support the findings of this study are available from the corresponding author upon reasonable request. 

\bibliography{bibliografia}

\clearpage
\onecolumngrid   
\section*{Supplemental Material}

\setcounter{figure}{0}
\setcounter{equation}{0}
\setcounter{table}{0}
\setcounter{section}{0}
\renewcommand{\thefigure}{S\arabic{figure}}
\renewcommand{\theequation}{S\arabic{equation}}

\newcommand{\IcRn}{I_cR_N}
\newcommand{\ETh}{E_{\mathrm{Th}}^{\mathrm{eff}}}
\renewcommand{\thefigure}{S\arabic{figure}}
\renewcommand{\thetable}{S\arabic{table}}
\renewcommand{\theequation}{S\arabic{equation}}
\title{Supplemental Material for\\
\emph{Transport regimes in tunnel ferromagnetic Josephson junctions}}
\author{}
\date{}

\maketitle

\section{Self-consistent description of the $s/F/S_R$ electrode}

The solid theory curves in Figs.~2(b) and 3(b) of the main text are obtained from the self-consistent calculation described here. The reduced nonlinear-tail model introduced in Sec.~\ref{sec:tailfit} is an independent description of Sample C and is not used to generate either main-text curve.

The right proximity electrode is represented as a one-dimensional $s/F/S_R$ structure. The spectral angle and the pair potentials in $s$ and $S_R$ are determined self-consistently at each temperature, and the two metallic interfaces obey the Kupriyanov--Lukichev boundary conditions \cite{Kurpianov1988_SN}. To zeroth order in the transparency of the AlO$_x$ barrier, the $I/s$ boundary is reflecting. A zero-gradient condition is imposed at the outer surface of the $S_R$ segment, whose length is $d_R=10\xi_{D,R}$.

The parameters are summarized in Table~\ref{tab:effective_sets}. Both Nb devices use the same superconducting scales, $T_c=8$~K and $T_{c,s}=6$~K, the same mismatch parameters $\gamma_{sF}=\gamma_{FS}=1$, and the common interlayer normalization $\xi_s\equiv\xi_{D,s}^{\rm eff}=10.7$~nm. Their nominal thicknesses then give $d_s/\xi_s=2.80$ for Sample B and $0.9346$ for Sample C. The temperature-independent factor $\zeta_I$ multiplies the tunnel current but does not alter the self-consistent spectral profiles or the $0$/$\pi$ branch. Its values, $0.73$ and $0.59$, differ by $19\%$. The Py and metallic-interface parameters are correlated effective quantities that characterize the processed multilayer rather than unique microscopic material constants.

\begin{table}[H]
\caption{Parameters of the self-consistent curves in Figs.~2(b) and 3(b) of the main text. Both samples use $T_c=8$~K, $T_{c,s}=6$~K, and $\gamma_{sF}=\gamma_{FS}=1$. The minimal symmetric parametrization $\gamma_B\equiv\gamma_{B,sF}=\gamma_{B,FS}$ refers to the two metallic interfaces and is distinct from the AlO$_x$ barrier. The listed RMSE values refer to the self-consistent curves. The current factor $\zeta_I$ is unrelated to the reduced-tail source factor $\eta_q$ in Sec.~\ref{sec:tailfit}.}
\label{tab:effective_sets}
\centering
\small
\begin{tabular}{lccccccc}
\toprule
Set & $T_{c,s}/T_c$ & $d_s/\xi_s$ & $d_F/\xi_{D,F}$ & $h^{\rm eff}$ & $\gamma_B$ & $\zeta_I$ & RMSE ($\mu$V)\\
\midrule
Sample B & 0.75 & 2.80 & 0.35 & 1.0 & 2 & 0.73 & 15.03\\
Sample C & 0.75 & 0.9346 & 0.50 & 6.8 & 4 & 0.59 & 2.47\\
\bottomrule
\end{tabular}
\end{table}

\section{Sample B: proximity-rounded crossover}
\begin{figure}[H]
\centering
\includegraphics[width=0.96\textwidth]{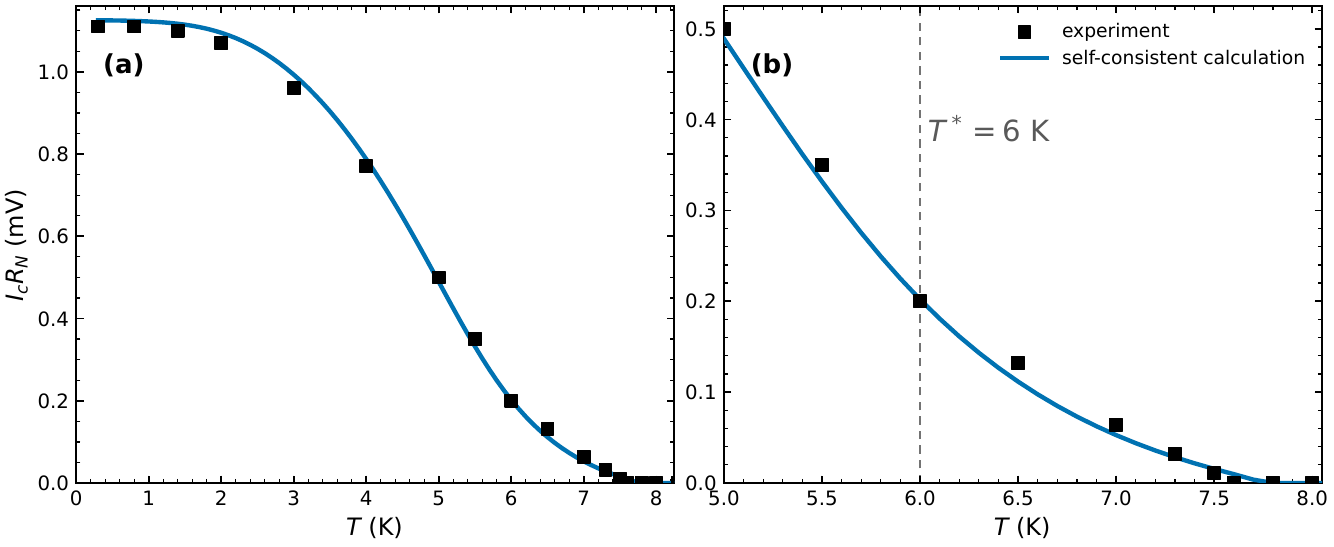}
\caption{Self-consistent result for Sample B. Squares are experimental values and the solid curve is the calculation. Panel (a) shows the full range and panel (b) resolves the high-temperature tail. The dashed line marks the intrinsic interlayer pairing scale $T^*=T_{c,s}=6$~K, not a separate transition of the coupled junction.}
\label{fig:S_Bfit}
\end{figure}
\begin{figure}[H]
\centering
\includegraphics[width=0.99\textwidth]{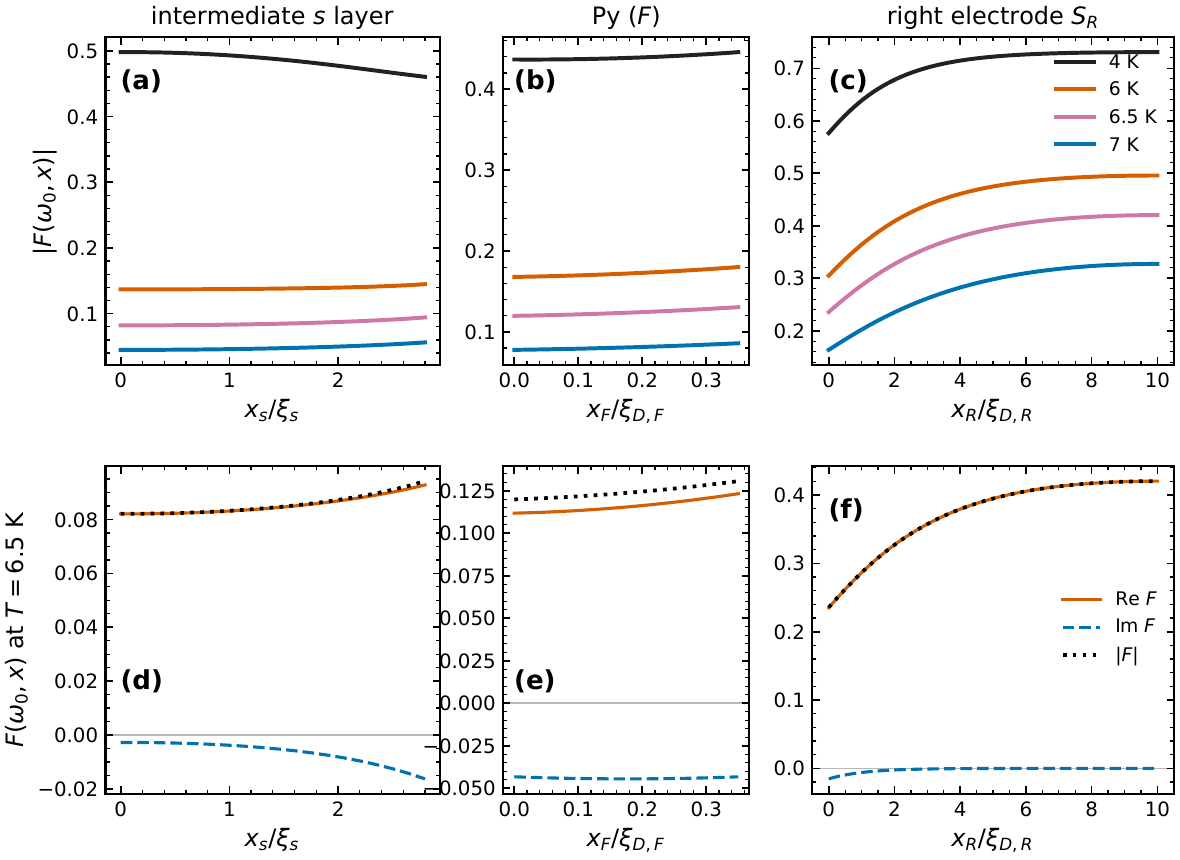}
\caption{Lowest-Matsubara anomalous amplitude for the same Sample-B solution used in Fig.~\ref{fig:S_Bfit}. Panels (a)--(c) show $|F(\omega_0,x)|$ in the intermediate $s$ layer, Py, and the right electrode $S_R$ at 4, 6, 6.5, and 7~K. The coordinate in each material starts at its left interface; $x_s=0$ is the tunnel side. Panels (d)--(f) show $\mathrm{Re}\,F$, $\mathrm{Im}\,F$, and $|F|$ at 6.5~K. The breaks between the three materials represent the finite interface jumps imposed by the Kupriyanov--Lukichev conditions.}
\label{fig:S_BFprofiles}
\end{figure}

For Sample B, the nominal 30-nm interlayer and common effective length $\xi_s=10.7$~nm give $d_s/\xi_s=2.80$. The intrinsic pairing scale of the interlayer is $T_{c,s}=T^*=6.0$~K. The remaining parameters and the tunnel-current normalization $\zeta_I=0.73$ are listed in Table~\ref{tab:effective_sets}. The curve reproduces the low-temperature amplitude and the finite-current tail above the intrinsic interlayer scale $T^*=6$~K. The origin of this tail is visible in the lowest-Matsubara anomalous amplitude
\begin{equation}
F(\omega_0,x)=\sin\theta(\omega_0,x),
\qquad \omega_0=\pi k_BT.
\end{equation}
Figure~\ref{fig:S_BFprofiles} follows this amplitude from the tunnel side of the intermediate layer, across Py, and into $S_R$. At and above $T^*$, correlations transmitted from $S_R$ through Py remain finite throughout the 30-nm $s$ layer and reach the tunnel boundary.
\begin{figure}[H]
\centering
\includegraphics[width=0.90\textwidth]{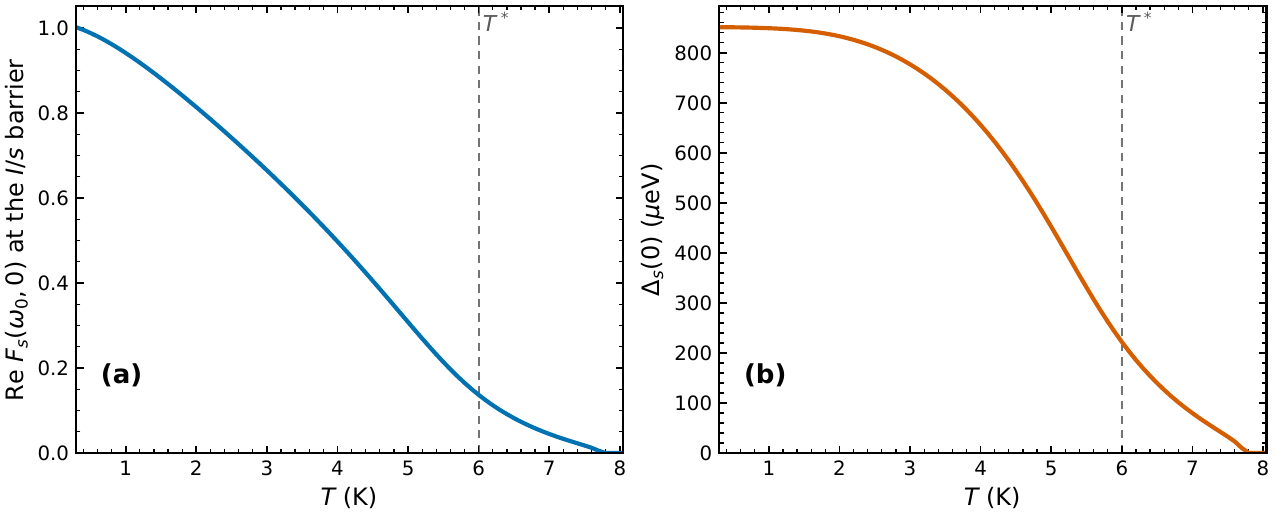}
\caption{Tunnel-side amplitudes for the Sample-B solution. (a) Real part of the lowest-Matsubara anomalous amplitude at the $I/s$ boundary. (b) Self-consistent interlayer pair potential at the same boundary. The vertical dashed line marks the intrinsic pairing scale $T^*=6$~K. Both quantities remain finite above $T^*$ because of coupling to $S_R$ through Py.}
\label{fig:S_Btunnel}
\end{figure}
The tunnel-side value provides a compact measure of the correlations available to carry current through the AlO$_x$ barrier. Neither $F_s(\omega_0,0)$ nor the self-consistent $\Delta_s(0)$ vanishes at $T^*$ in the coupled structure. Instead, $T^*$ marks a proximity-rounded crossover from predominantly intrinsic interlayer superconductivity to a regime dominated by correlations supplied from $S_R$. The role of Py is therefore twofold: its exchange field suppresses the adjacent condensate through inverse proximity, while the same layer transmits correlations from $S_R$. Within the symmetric interface parametrization, $\gamma_B=2$ limits both processes. Their balance produces the rounded crossover at $T^*$ rather than a separate transition of the coupled device. The solution remains on the $0$ branch throughout the measured range.  
\section{Sample C: proximity-supported suppressed-$s$ regime}

For Sample C we retain the same $T_c=8.0$~K electrode scale, $T_{c,s}=6.0$~K intrinsic interlayer pairing scale, and $\xi_s=10.7$~nm effective length as for Sample B. The nominal 10-nm interlayer therefore gives $d_s/\xi_s=0.9346$. Here $T_{c,s}$ is the pairing scale entering the self-consistency equation, not the transition temperature of the coupled layer; the remaining parameters are listed in Table~\ref{tab:effective_sets}.
The current normalization $\zeta_I=0.59$ gives an RMSE of $2.47~\mu$V over $0.5$--$4$ K. The 10-mK point in Fig.~3(b) of the main text is displayed but is not included in this fit. Within the symmetric effective-interface parametrization, the larger value $\gamma_B=4$ weakens both reverse-proximity pair breaking of the thin $s$ layer and transmission of correlations from $S_R$. It is therefore part of a correlated description of the processed interfaces rather than an independent measurement of either interface resistance.

The self-consistent interlayer pair potential remains strongly suppressed. Its largest absolute value in the $0.5$-K profile is approximately $33~\mu$eV, much smaller than the meV-scale pair potential in the right Nb electrode.

\begin{figure}[H]
\centering
\includegraphics[width=0.82\textwidth]{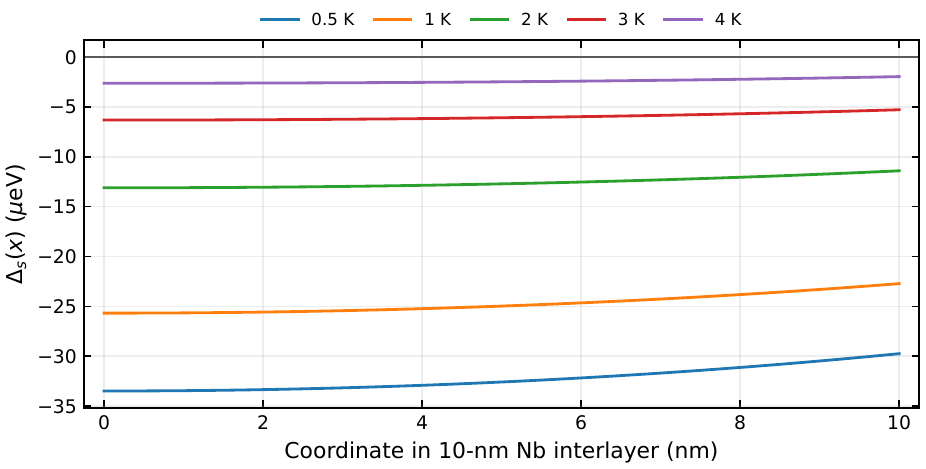}
\caption{Self-consistent pair-potential profiles in the 10-nm Nb interlayer for $h_C^{\rm eff}=6.8$ and $\gamma_B=4$ at $T=0.5$, $1$, $2$, $3$, and $4$ K. In the convention where the pair potential in the right Nb electrode is positive, the negative interlayer pair potential identifies the $\pi$ branch. Its maximum magnitude is approximately $33~\mu$eV at $0.5$ K.}
\label{fig:S_Deltas}
\end{figure}
In the phase convention used here, $\Delta_R>0$ in the right Nb electrode, whereas $\Delta_s<0$ and the signed first Josephson harmonic is negative. The Sample-C solution therefore lies on the $\pi$ branch throughout the measured range. This is a model prediction: a single-junction current--voltage characteristic determines $I_c=|I_1|$, but not the sign of $I_1$. Verification requires a phase-sensitive measurement.

\begin{figure}[H]
\centering
\includegraphics[width=0.82\textwidth]{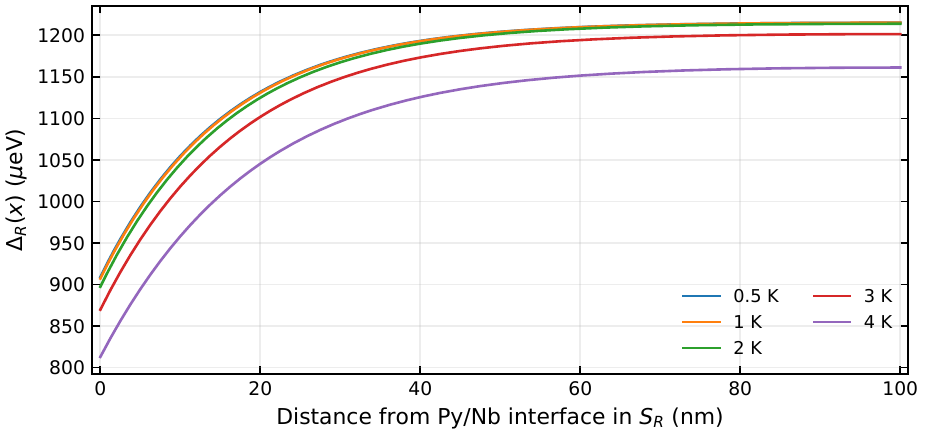}
\caption{Right-electrode profiles from the same Sample-C solution as Fig.~\ref{fig:S_Deltas}. The coordinate is measured from the Py/Nb interface. The order parameter recovers monotonically toward its bulk value away from Py.}
\label{fig:S_DeltaR}
\end{figure}

\begin{figure}[H]
\centering
\includegraphics[width=0.99\textwidth]{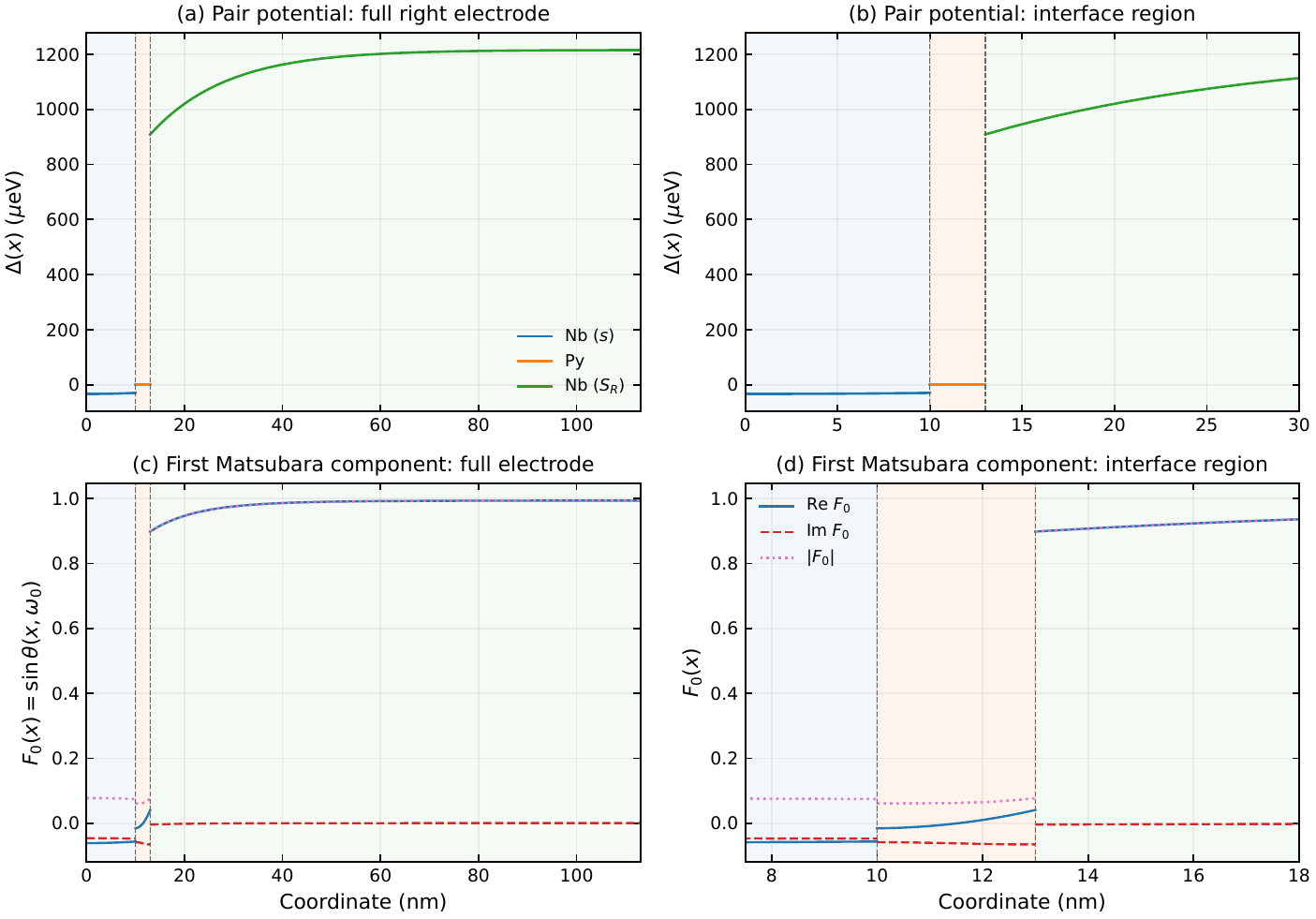}
\caption{Pair potential and lowest-Matsubara anomalous amplitude across the Sample-C $s/F/S_R$ electrode at $T=0.5$ K. Panels (a,b) show $\Delta(x)$ on the full scale and in the interface region; panels (c,d) show $\mathrm{Re}\,F_0$, $\mathrm{Im}\,F_0$, and $|F_0|$. The finite, generally unequal interface breaks follow from the Kupriyanov--Lukichev conditions and the different local spectral amplitudes. The pair potential in $s$ is strongly suppressed, while the tunnel-side anomalous amplitude remains finite.}
\label{fig:S_F0Delta}
\end{figure}

The 10-nm layer is therefore not a robust independent superconducting electrode. Its finite tunnel-side anomalous amplitude reflects both the small local pair potential and correlations transmitted from $S_R$ through Py. The tunnel barrier consequently couples $S_L$ to the composite $s/F/S_R$ electrode. Figure~\ref{fig:S_F0Delta} compares the pair potential with the lowest positive-Matsubara component
\begin{equation}
F_0(x)=\sin\theta(x,\omega_0),
\qquad \omega_0=\pi k_BT.
\end{equation}
At $T=0.5$ K, $\Delta_s$ is strongly suppressed, whereas $|F_0|$ remains finite on the tunnel side. Sample C is therefore close to the SInFS limit but retains a small nonzero self-consistent pair potential.

\section{Reduced description of Sample C}\label{sec:tailfit}

In the strongly suppressed-$s$ limit, the temperature dependence can also be described by the nonlinear long-junction tail generated by the right source electrode. Using the $8$-K electrode scale of the main text, we write
\begin{align}
q_n&=\eta_q\tan\frac{\theta_R(\omega_n)}{4},\\
\Lambda_n&=\sqrt{\frac{2\omega_n}{\ETh}},\\
u_n&=2q_n e^{-\Lambda_n},
\end{align}
where $\theta_R(\omega_n)=\arctan[\Delta_R^{\rm bulk}(T)/\omega_n]$ is the bulk spectral angle of the right source electrode. The tunnel-side anomalous amplitude is
\begin{equation}
\sin\theta(0,\omega_n)=\frac{4u_n(1-u_n^2)}{(1+u_n^2)^2}.
\end{equation}

\begin{figure}[H]
\centering
\includegraphics[width=0.78\textwidth]{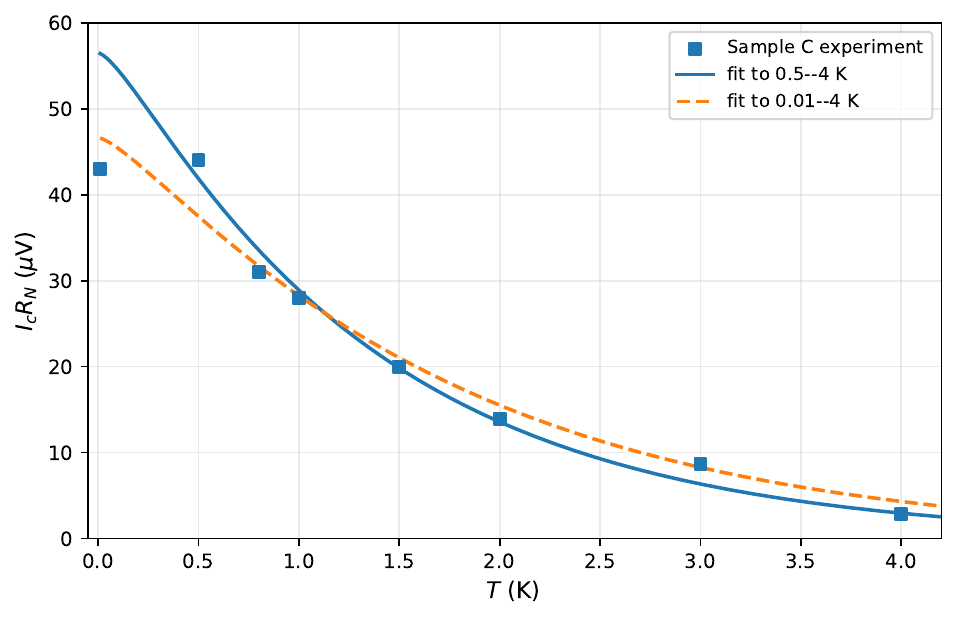}
\caption{Independent reduced-tail fits for Sample C. Squares are the experimental data. The solid curve is fitted over $0.5$--$4$ K, with the 10-mK point shown but excluded; the dashed curve includes the 10-mK point. These curves are distinct from the self-consistent result in Fig.~3(b) of the main text and do not determine its parameters.}
\label{fig:S_fit_ranges}
\end{figure}

The corresponding dimensionless propagation length is
\begin{equation}
L_{\rm fit}=\sqrt{\frac{2\pi k_BT_c}{E_{\mathrm{Th}}^{\mathrm{eff}}}},
\end{equation}
in units of the thermal diffusion length associated with the 8-K electrode scale.

\begin{table}[H]
\caption{Parameters of the independent reduced-tail fits in Fig.~\ref{fig:S_fit_ranges}.}
\label{tab:tail_parameters}
\centering
\begin{tabular}{lcccc}
\toprule
Fit range & $\ETh$ ($\mu$eV) & $\eta_q$ & $L_{\rm fit}$ & RMSE ($\mu$V)\\
\midrule
$0.5$--$4$ K & 71.5 & 0.278 & 7.79 & 1.59\\
$0.01$--$4$ K & 101.1 & 0.169 & 6.55 & 2.80\\
\bottomrule
\end{tabular}
\end{table}

The fitted $\ETh$ is a composite spectral scale containing propagation through the suppressed Nb layer, damping and phase rotation in Py, and finite-interface suppression; it is not the bare Thouless energy of an ideal transparent SNS junction. Boundary-induced reductions of effective Thouless scales in related SN--N--NS structures are discussed in Ref.~\cite{Bakurskiy2024_Thouless}.

\section{Length and exchange-energy scales}

For the Nb electrodes, the Ginzburg--Landau coherence length at zero temperature can be extracted from the linear dependence of the perpendicular upper critical field,
\begin{equation}
H_{\mathrm{c2}\perp}(T)=\frac{\Phi_0}{2\pi\xi_{\rm GL}^2(0)}
\left(1-\frac{T}{T_c}\right).
\end{equation}
Using the $90\%$ normal-resistance criterion gives $\xi_{\rm GL}(0)\simeq10$~nm for the 400-nm Nb electrode film. This experimental scale is close to, but not identified with, the dirty-limit normalization of the one-dimensional model. For the two Nb interlayers we use the common effective length defined by the Sample-B thickness ratio,
\begin{equation}
\xi_s\equiv\xi_{D,s}^{\rm eff}=10.7~\mathrm{nm},
\qquad \frac{30~\mathrm{nm}}{\xi_s}=2.80,
\qquad \frac{10~\mathrm{nm}}{\xi_s}=0.9346.
\end{equation}
The nominal 10-nm Sample-C layer then has $d_s/\xi_s=0.9346$. In both samples, $T_{c,s}=6$~K denotes the intrinsic pairing scale entering the self-consistency equation, not the transition temperature of the proximity-coupled layer.

The thermal diffusion normalization length and magnetic decay length are distinct:
\begin{align}
\xi_{D,F}&=\sqrt{\frac{\hbar D_F}{2\pi k_BT_c}},\\
\xi_{F1}&=\xi_{D,F}\sqrt{\frac{2}{h}}=\sqrt{\frac{\hbar D_F}{H}}.
\end{align}
The 3-nm Py resistivity is $\rho_{\rm Py}=84~\mu\Omega\,\mathrm{cm}$. Using $D_F=v_Fl_F/3$, $v_F=2.2\times10^5$~m/s, and $\rho l_F=31.5\times10^{-6}~\mu\Omega\,\mathrm{cm}^2$ \cite{Mayadas1974} gives $D_F\simeq2.75$ cm$^2$/s and $\xi_{D,F}\simeq6.5$ nm at $8$ K. With the bare Py exchange energy $H\simeq200$ meV \cite{Robinson2006}, the magnetic decay length is $\xi_{F1}\simeq0.95$ nm. The calculation uses $\xi_{D,F}$ to nondimensionalize the coordinate and retains the exchange field explicitly through $h$.

For a 3-nm Py film these material estimates give $d_F/\xi_{D,F}\simeq0.46$ and $h_{\rm bare}\simeq92$. In the one-dimensional description, however, the processed ultrathin Py layer and its two interfaces are represented collectively by $(d_F/\xi_{D,F},h^{\rm eff},\gamma_B)=(0.35,1.0,2)$ for Sample B and $(0.50,6.8,4)$ for Sample C. These correlated effective parameters characterize the integrated spectral damping, phase rotation, and interface suppression; they are not direct determinations of the bare Py exchange energy.
\end{document}